# First-principles prediction of altermagnetism in transition metal graphite intercalation compounds

Xiaojun He,[a‡] Guo-Dong Zhao, [b‡] Weida Fu [a], Tao Hu,[a*] Wencai Yi[d], Hui Zhang[b], Alessandro Stroppa[c], Wei Ren[b] and Zhongming Ren[a]

[a]School of Materials Science and Engineering, State Key laboratory of Advanced Special Steel, Shanghai University, Shanghai, 200241, China
[b]Department of Physics, Materials Genome Institute, International Centre for Quantum and Molecular Structures, Shanghai University, Shanghai 200444, China
[c]CNR-SPIN c/o Università degli Studi dell'Aquila, Via Vetoio 10, I-67010 Coppito (L'Aquila), Italy
[d]Laboratory of High Pressure Physics and Material Science, School of Physics and Physical Engineering, Qufu Normal University, Qufu 273165, China

We report the emergence of altermagnetism, a magnetic phase characterized by the coexistence of compensated spin ordering and momentum-dependent spin splitting, in graphite intercalation compounds (GICs), a prototypical material system long investigated for its tunable electronic and structural properties. Through first-principles calculations, we demonstrate that vanadium-intercalated stage-1 graphite compounds, exhibit inherent altermagnetic properties. The hexagonal crystal and antiferromagnetic ordering of V atoms generate a spin group ($P^{-1}6_3^1m^{-1}c^{\infty m}1$) that enforces alternating spin polarization in momentum space while maintaining zero net magnetization. The calculated band structure reveals robust altermagnetic signatures: along the $M_1' - \Gamma - M_1$ high-symmetry direction, we observe a pronounced spin splitting of ~272 meV with alternating spin polarization. Crucially, the spin splitting exhibits minimal sensitivity to spin-orbit coupling (SOC) effect, highlighting the dominance of exchange interactions over relativistic effects. From Monte Carlo simulations, we predict a magnetic transition temperature ($T_m$) of ~ 430 K. This well-above-room-temperature value positions the material as a promising candidate for application-oriented altermagnetic research. These findings position V-GICs as a promising high-temperature altermagnet, ideal for zero-field spin-polarized current generation and topologically protected spin transport. As the first demonstration of a metallic carbon-based altermagnet, our work establishes a promising platform for developing next-generation carbon-based electronic materials functional at room temperature.

## INTRODUCTION

The recent discovery of altermagnetism [1–7], a collinear magnetic phase characterized by non-trivial spin group symmetry, has expanded the conceptual framework of quantum magnetism while offering considerable potential for spintronic applications [8–13]. This emergent phase combines three defining characteristics: robust time-reversal symmetry breaking enabling spin-polarized electronic states, antiparallel magnetic ordering maintaining zero net magnetization, and momentum-dependent alternating spin-split band structures [14–20]. Unlike conventional ferromagnetism or antiferromagnetism, altermagnetism arises from specific crystallographic constraints, being strictly confined to 37 spin point groups where sublattice symmetry operations form index-2 subgroups of the parent crystal structure [21–28]. Such symmetry-engineered hybridization creates unique quantum phenomena previously considered mutually exclusive in traditional magnetic classifications, as evidenced by experimental confirmations of theoretically predicted signatures including anomalous Hall/Nernst effects, nonlinear spin currents, and magneto-optical responses [29–32]. While these breakthroughs validate the unique symmetry-broken states and device potential of altermagnetism, which has been identified in numerous metallic compounds [33,34], no verified altermagnetic phases have been reported in carbon-based materials. This challenge motivates re-examination of established material platforms through the lens of symmetry-controlled magnetism.

Graphite intercalation compounds (GICs) is a traditional material system that has been extensively studied, with leading researchers Dresselhaus uncovering a vast range of fascinating properties [35–37]. From a symmetric perspective, intercalated graphite has great potential as a platform for discovering and harnessing unique magnetic phenomena like altermagnetism. The layered structure of graphite allows for the controlled insertion of transition metals [38] which can introduce local magnetic moments and disrupt the symmetry of the system. This symmetry breaking, coupled with the interaction between the intercalant's *d*-electrons and the π-electrons of graphite [39,40], can give rise to complex magnetic orders [41], potentially including altermagnetism. Therefore, the tunable electronic and magnetic properties of intercalated graphite make it a promising candidate for studying the emergence of altermagnetism and its associated phenomena.

In this work, we systematically investigate the emergence of altermagnetism in transition metal-intercalated graphite from a symmetry perspective, employing first-principles calculations to validate this novel magnetic phase. Our calculations identify vanadium graphite intercalation compounds (V-GIC) as a robust altermagnetic host, exhibiting momentum-dependent spin-split band structures along specific high-symmetry paths. Crucially, these spin-split bands demonstrate minimal sensitivity to spin-orbit coupling (SOC) effects, with splitting energies governed primarily by exchange interactions. The V-GIC exhibits a magnetic anisotropy energy (MAE) of 0.6 meV per formula unit and stabilizes in-plane magnetic moments, as revealed by relativistic calculations. Monte Carlo simulations further predict a magnetic transition temperature ($T_m$) of ~430 K, demonstrating robust magnetic ordering stability within the operational temperature range for practical applications. Our study not only unravels the fundamental mechanisms driving altermagnetism in layered intercalation compounds but also provides a framework for developing next-generation carbon-based electronic materials. The subsequent sections detail our computational methodology, present comprehensive results across atomic structure, electronic properties, and magnetic, and provide comparative analyses with conventional magnetic systems.

## METHODS

The density functional theory (DFT) calculations described in this paper were implemented in the Vienna ab initio simulation package (VASP) code [42–44]. The atomic-scale simulation models were developed using Device Studio, a comprehensive computational package created by Hongzhiwei Technology (Shanghai) Co., Ltd. DFT computations were partially conducted through the integrated Projector-Augmented Wave (PAW) method implementation within the DS-PAW software suite. PAW potentials were employed to describe the interaction between valence electrons and ion core [45,46], ensuring accurate representation of the electronic structure. A plane-wave basis set with a kinetic energy cutoff of 500 eV was utilized, and the Brillouin zone was sampled using a 9×9×4 Γ-centered Monkhorst-Pack *k*-grid for structure relaxation calculations. To account for the correlation energy of the strongly localized 3*d* orbitals of transition metals (TM), the Hubbard U correction (DFT+U) was applied [47,48], with the Dudarev type [49,50], utilizing the exchange-correlation functional of Perdew-Burke-Ernzerhof (PBE) [51] type in generalized gradient approximation (GGA). Previous studies of TM-adsorbed graphene systems using GGA+U have shown consistency with results obtained from the B3LYP hybrid functional, providing a robust validation of our approach [52]. All structures were fully relaxed without imposing symmetry constraints, with convergence criteria of $10^{-7}$ eV for energy and 0.01 eV/Å for Hellmann-Feynman forces. The intercalation energy ($E_{ic}$) was defined as $E_{ic} = E_{VGIC} - E_V - E_G$, $E_{VGIC}$, $E_V$ and $E_G$ are the energies of the intercalated compounds $C_{16}V_2$, V atoms and bulk graphite. It is essentially that the formation energy of the compound is used to assess the thermodynamic stability of the system [53,54]. To investigate the dynamic and thermal stabilities, phonon spectra are simulated with the density functional perturbation theory (DFPT) by the Phonopy code [55], and the *ab initio* molecular dynamics (AIMD), the cutoff energy of 400 eV was set and the energy convergence criterion was set to $2 \times 10^{-5}$ eV. As supercell contains nearly 486 atoms, only the Γ-point sampling was used to sample the Brillouin zone. Molecular dynamics simulations were carried out using the canonical (NVT) ensemble at a temperature of 300 K for a duration of 10 ps with a time step of 2 fs. The Nosé-Hoover thermostat was utilized to maintain the system temperature, as its stochastic nature is well-suited for phase space sampling. The spin-orbit coupling (SOC) is considered for different spin orientations in the magnetic anisotropy energy (MAE) calculations with the energy convergence criteria of $10^{-8}$ eV as MAE = E[001] − E[100]. The SOC is also included to obtain electronic structure when it is necessary. Classical Metropolis Monte Carlo (MC) simulations of the Heisenberg model [56,57] were performed to calculate the magnetic phase transition temperatures with 10 × 10 supercell, 6500 temperature points. On each temperature point, $10^5$ Monte Carlo steps were performed for equilibrium and another $10^5$ steps were performed for statistics [58]. The specific heat $C_v = \frac{\langle E^2\rangle - \langle E^2\rangle}{N k_B T^2}$ where E is the total energy, N is the number of magnetic ions, $k_B$ is the Boltzmann constant, and $T$ is the temperature.

## RESULTS and DISCUSSION

The core challenge in achieving altermagnetism lies in the precise design of symmetry breaking, particularly at the crystal structure level. This challenge can be summarized as follows. First, the establishment of a magnetic ordered state necessitates the breaking of time-reversal symmetry (T), which is a prerequisite for the generation of spin-polarized electronic bands. While breaking time-reversal symmetry, it is crucial to maintain the joint operation formed by both the magnetic order and crystal symmetry, such as $T \cdot C_n$ (time-reversal coupled with n-fold rotation) or $T \cdot M$ (time-reversal combined with mirror reflection), which ensures the alternating distribution of spin polarization in momentum space, ultimately achieving zero net magnetization.

Altermagnetism can be observed in stage-1 intercalated graphite, which can be understood as a bilayer system composed of two graphite layers and two single ferromagnetic layers. Within this system, magnetic metal atoms are stacked through antiferromagnetic coupling, and the graphite layers are magnetized due to their different spatial positions. Essentially, it meets the requirement that opposite sublattices cannot be interchanged by simple inversion or translation. The archetypal system of intercalated graphite precisely meets these requirements. Specifically, the $C_{16}X_2$ compound investigated here, a graphite intercalation compound with transition metals, exhibits key structural prerequisites: it crystallizes in a the non-centrostmmetric space group $P6_3mc$ (No.186). This asymmetry condition alone, however, does not guarantee actual altermagnetic ordering. Through first-principles calculations of the magnetic ground state of various transition metal intercalated graphite, we found that only V-GIC perfectly simultaneously satisfies the above conditions, whereas other intercalated compounds exhibit exclusively interlayer ferromagnetic coupling, as detailed in Table 1. Furthermore, by utilizing the AMCHECK package [59], we confirmed that the system fulfills the criteria for altermagnetic ordering when its net magnetic moment vanishes.

Table 1. Magnetic ground states of transition metal-intercalated graphite compounds. AFM and FM denote antiferromagnetic and ferromagnetic ground states in transition metal-intercalated graphite compounds, respectively. ΔE denotes the energy difference between AFM and FM states per formula units. The magnetic moments listed are those of the transition metal atoms.

| TM | V | Cr | Mn | Fe | Co |
|---|---|---|---|---|---|
| Magnetic Ground State | AFM | FM | FM | FM | FM |
| ΔE (eV) | 0.251 | 0.401 | 0.703 | 0.153 | 0.125 |
| $U_{eff}$ (eV) | 5.0 | 3.4 | 4.8 | 5.1 | 6.2 |
| Magnetic Moment ($\mu B$) | 2.63 | 3.51 | 3.10 | 2.17 | 1.77 |

We characterized the crystal structure of vanadium graphite intercalation compounds revealing a stage-1 configuration with unit cell composition $C_{16}V_2$. The compound crystallizes in a hexagonal system (space group $P6_3mc$, No.186 and 6mm point group) with lattice parameters $a = b = 5.011$ Å and $c = 7.286$ Å, and unit cell angles $\alpha = \beta = 90°$ and $\gamma = 120°$, consistent with layered intercalation symmetry. The symmetry operations possess $6_3$ screw axes along [001], forming a polar structure lacking inversion symmetry. Vanadium atoms occupy the 2b Wyckoff positions (multiplicity 2) between carbon layers, where equivalent positions are generated through $6_3$ screw operations necessitates a spin-flip, enforcing antiparallel spin alignment in the in-plane direction. Carbon sublattices maintain hexagonal symmetry through 6c Wyckoff sites (multiplicity 6) with interlayer zero net magnetization, as shown in Fig. 1. This magnetic configuration preserves the parent graphite's hexagonal framework while introducing ordered V intercalates, where the altermagnetic order emerges from the interplay between $6_3$ screw-axis mediated structural symmetry and antiferromagnetic exchange coupling.

To investigate the structural stability of V-GIC, we performed AIMD simulations on the supercell structure at room temperature. Under thermal perturbations, all atoms exhibited regular thermal vibrations while maintaining structural integrity without significant distortion, with the final configuration visualized in Figure 1(c). The time evolution profiles of total energy and temperature (Fig. 1d) demonstrate excellent convergence throughout the simulation. The temperature oscillates within 300 ± 30 K while the total energy variation remains below 0.001 eV/atom.

These computational results collectively confirm that the system exhibits excellent structural stability under ambient conditions. The intercalated energy ($E_{ic}$) is -0.93 eV per V atom, where a negative value indicates an exothermic intercalated, revealing that the V-GIC is thermodynamically stable. The dynamic stability of the structure was confirmed by phonon spectrum analysis in Figure 1(e), and no obvious imaginary frequencies were observed

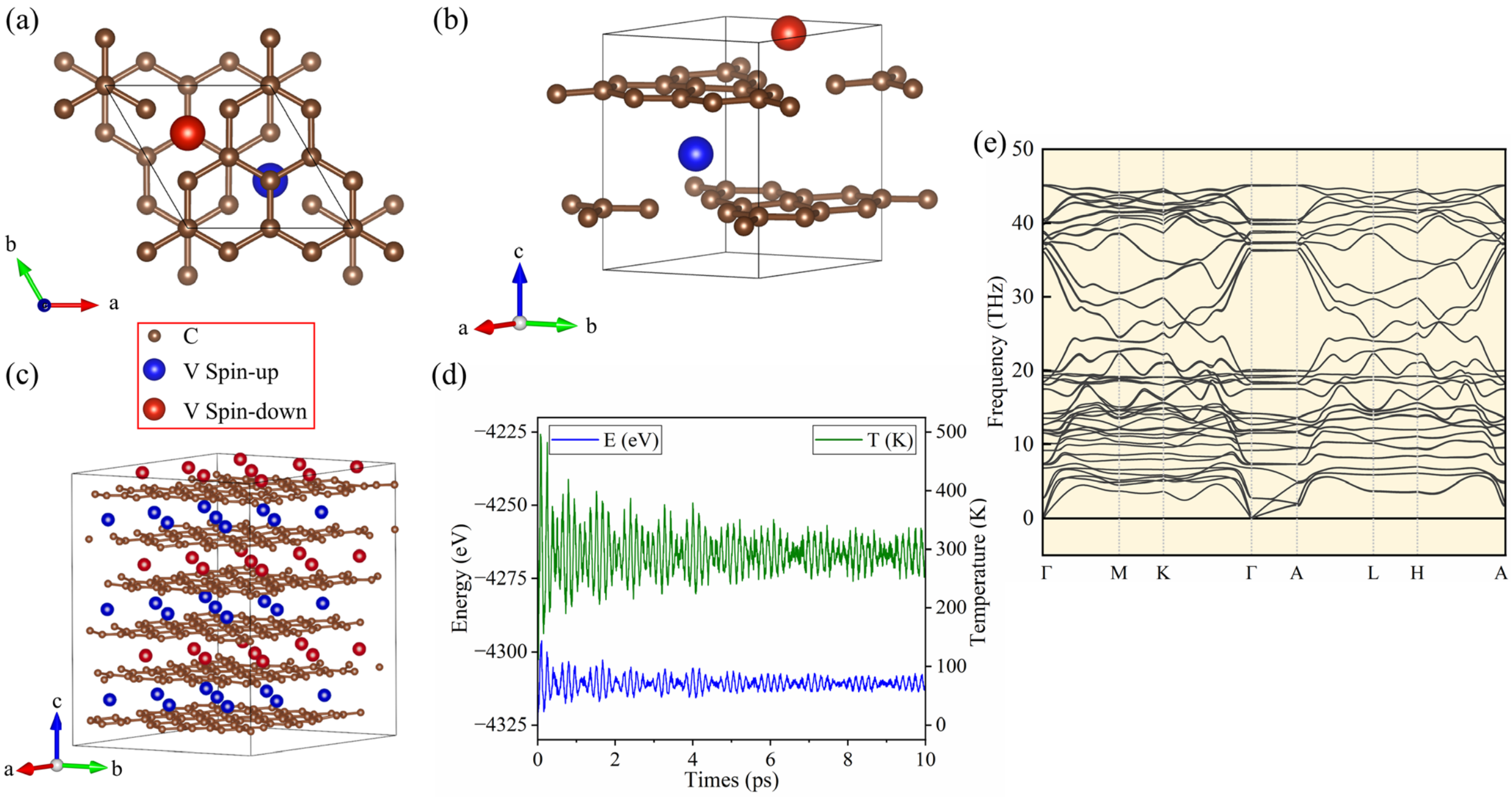

Fig 1. (a) The top view and (b) The Cross-sectional view of the V-GIC structure. Red arrows indicate the magnitude and orientation of the magnetic moments, the atomic magnetic moments of vanadium are depicted at their actual scale, whereas those of carbon atoms are artificially enhanced twentyfold for enhanced visual clarity. (c) Perspective view of the 3×3×3 supercell structure for vanadium-intercalated graphite after 10 ps AIMD simulation at 300 K, demonstrating structural relaxation to thermodynamic equilibrium. (d) Time evolution profiles of temperature fluctuation (green curve) and energy convergence (blue curve) during MD simulation. (e) Phonon spectra of V-GIC representing dynamic stabilities.

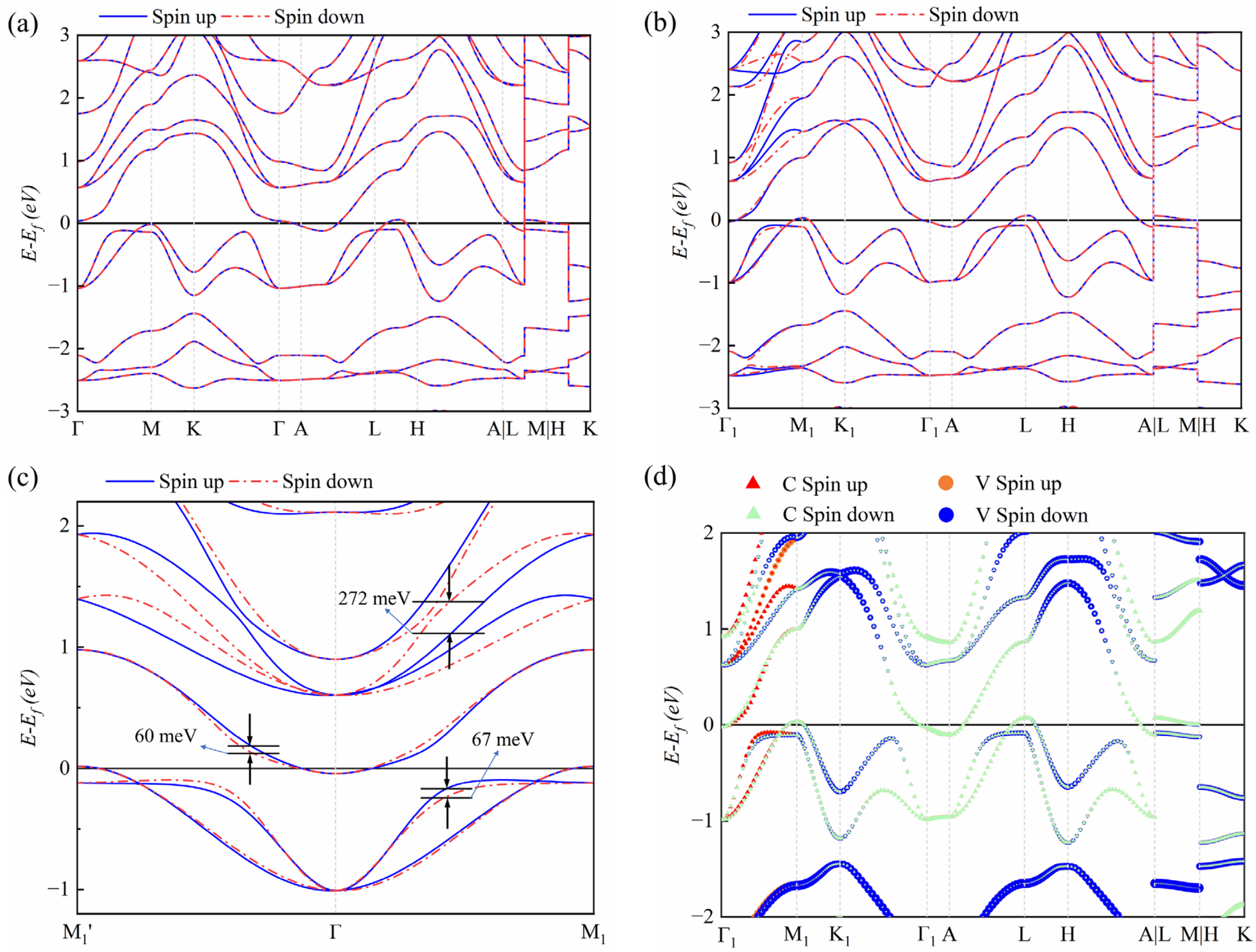

Fig 2. (a) Calculated spin-polarized band structure of the V-GIC along the conventional high-symmetry path, illustrating the energy dispersion of spin-up and spin-down states across the Brillouin zone. (b) Calculated band structure along a specific path including an off-nodal plane trajectory ($\Gamma_1 \rightarrow M_1 \rightarrow K_1$). The special points $\Gamma_1$, $M_1$, and $K_1$ are defined as the midpoints between the high-symmetry points Γ-A, M-L, and K-H, respectively. Panel (c) highlights the splitting along the $M_1'$-Γ-$M_1$ direction, the calculated band dispersion exhibits alternating spin polarization across the gamma point, some of the band splitting values are also labelled. (d) The fat energy band diagram of the density of states contributions, where the different color shapes of the bubbles represent the contributions of the different elements and spin channels, respectively. The size of the bubble indicates the weight of its contribution.

The spin-resolved band structure of the V-GIC system provides a key insight for understanding its alternating magnetic properties. As shown in Fig. 2a, the spin-up and spin-down bands exhibit perfectly overlapping alone conventional high-symmetry paths (same as graphite as shown in SM) in the Brillouin zone (BZ), forming nodal lines with vanishing net magnetization. However, symmetry analysis reveals that the system breaks combined inversion and time-reversal ($\mathcal{PT}$), preventing persistent spin degeneracy throughout the entire BZ, as evidenced by the band splitting along the $\Gamma \rightarrow M_1$ direction (Fig. 2b). This originates from connection between two sublattices in real space are not simple glide symmetries, but rather $C_{6z}|t$, $C_{2z}|t$ and $M|t$. Here, $C_{6z}|t$ denotes a six-fold rotation along z-axis combined with translation $t$, $C_{2z}|t$ similarly indicates a two-fold rotation plus $t$. $M$ represents mirror mappings along the (120), (210) and $(1\bar{1}0)$ planes, detailed in SM. To interpret this unconventional band topology, we further employ the non-relativistic spin group formalism. The nontrivial spin group is typically represented as $[R_i||R_j]$, where the left $R_i$ acts on spin space and the right $R_j$ acts on real space [25,60]. Therefore, we focus on $[C_2||R]$, $C_2$ on the left of the double vertical bar is a 180° spin-space rotation transformation around an axis perpendicular to the spin direction, while on the right bar R denotes the aforementioned operation connecting the two spin sublattices. Consequently,

$$\epsilon(s, \mathbf{k}) = [C_2||R]\epsilon(s, \mathbf{k}) = \epsilon(-s, R\mathbf{k}).$$

When $k_z = \frac{n\pi}{c}$ (n is an integer number), $C_{2z}tk$=-$k$. Considering an inversion,

$$[\bar{C}_2\|\mathcal{T}][C_2\|C_{2z}|t]\epsilon(s,\mathbf{k}) = [\bar{C}_2\|\mathcal{T}]\epsilon(-s,-\mathbf{k}) = \epsilon(-s,\mathbf{k}),$$

enforcing degeneracy in $k_x$-$k_y$ projection planes (blue planes in Fig. S2). Simultaneously, the green planes are protected by $[C_2\|M|t]$, leading to zero spin polarization ($\boldsymbol{P}_s$=0) and forming four spin-degenerate nodal planes cross the Γ point. Away from those symmetric regions, $\boldsymbol{P}_s \neq 0$, defining AM. Furthermore, we delve into the krammer pairs along the splitting path Γ→M₁. $C_{6z}t$ here satisfies $C_{6z}t\mathbf{k}(\Gamma - M_1) = \mathbf{k}(\Gamma - M_1')$, thus $\epsilon(s,\Gamma - M_1) = \epsilon(-s,\Gamma - M_1')$, yielding alternating $\boldsymbol{P}_s$ with a *g*-wave character as shown in Fig. 2c. Specifically, the maximum spin-splitting energy gap reaches approximately 272 meV. Near the Fermi energy level, its conduction and valence bands exhibit band splitting of 60 and 67 meV, respectively. This coexistence of alternating spin polarization and nodal planes degeneracy directly reflects the spin-momentum locking mechanism protected by symmetry, underscores the AM character. From the band structure, we can also see that V-GIC are clearly metallic alternating magnetic materials. Therefore, V-GIC can be employed as promising spin current controllers due to the advantage of metal's good conductivity and controllability. In particular, metal materials are more feasible and efficient in controlling spin currents through electric fields [61]

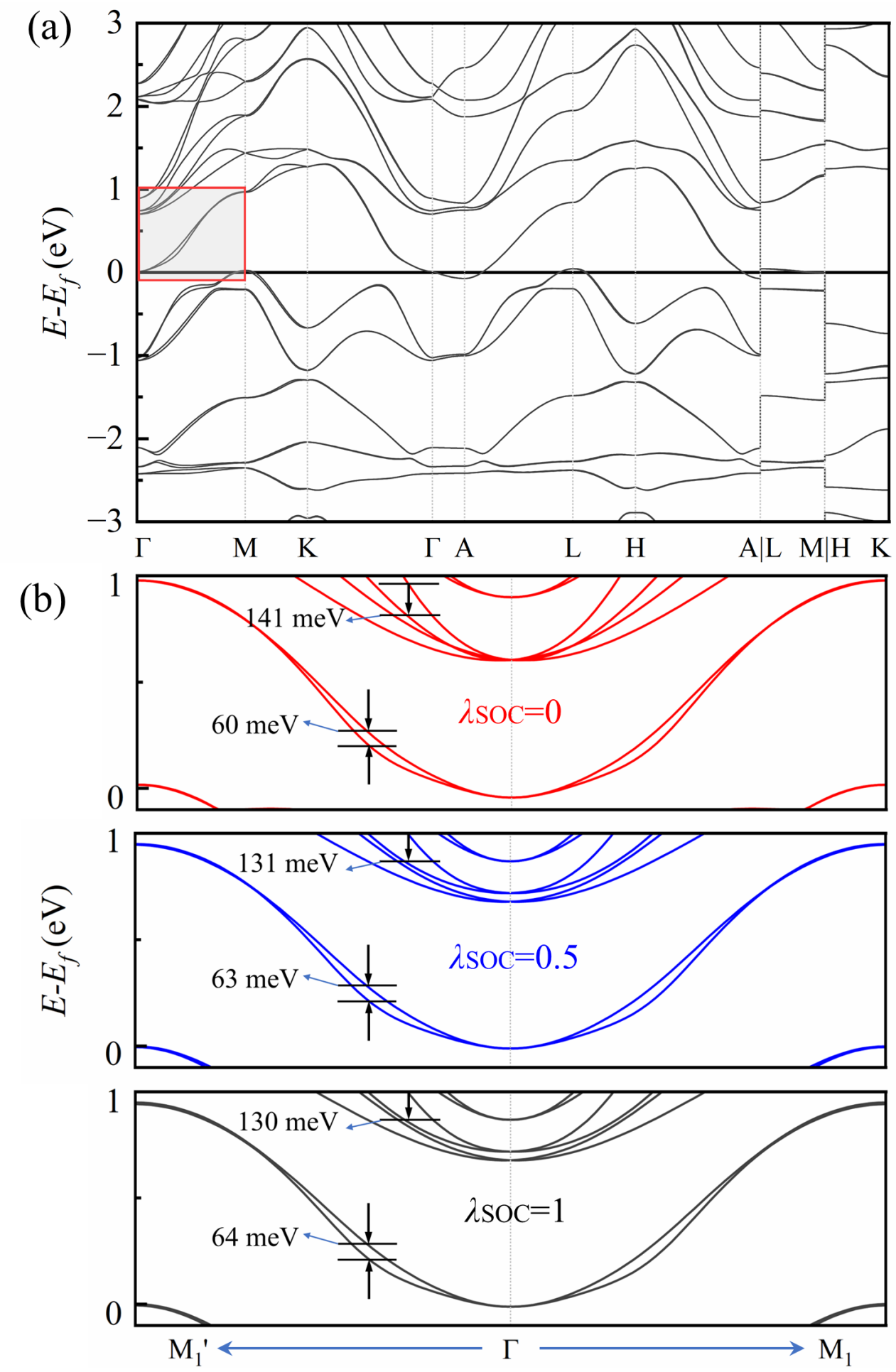

Fig 3. (a) The band structure including spin-orbit coupling, (b) highlighting the splitting along the M₁'-Γ-M₁ direction.

To investigate the effect of SOC strength on band structures, we introduce a scaling factor $\lambda_{\mathrm{SOC}}$ to the SOC Hamiltonian term within the DFT framework [60]. The SOC Hamiltonian is expressed as:

$$\mathcal{H}_{\mathrm{SOC}} = \lambda_{\mathrm{SOC}}\left(\frac{\hbar^2}{2m_e^2c^2}\frac{K(r)}{r}\frac{dV(r)}{dr}\hat{L}\cdot\hat{S}\right) \quad (1)$$

where $\hat{L} = \hat{r} \times \hat{P}$ is the orbital angular momentum operator, $\hat{S}$ is the spin operator, $V(r)$ is the spherical part of the effective all-electron potential within the PAW sphere, and $K(r) = \left(1 - \frac{V(r)}{2m_e c^2}\right)^{-2}$. All Hamiltonians, wave functions, charge densities, and potential are calculated self-consistently. Fig. 3a presents the band structure incorporating SOC effects. Compared to the SOC-free case, a small band splitting modification is observed. Crucially, this SOC-induced splitting has a negligible impact on the overall band structure, particularly on the pre-existing split bands. This perturbative role is further quantitatively demonstrated in the magnified band structures in Fig. 3b, which compare the cases of $\lambda_{SOC}$ = 0, 0.5, and 1. A closer inspection to these SOC induced modulations provides a crucial insight into the origin of the splitting. For the bands along $M_1$'-Γ-$M_1$, the splitting enhanced from 60 meV to 64 meV while the splitting in the same momentum region is substantially reduced from 141 meV to 130 meV. This complex response to SOC further corroborates that it is not the primary driving mechanism. Therefore, the presence of a large, intrinsic splitting that is merely perturbed by SOC provides compelling evidence that the spin-polarized band structure is a hallmark of AM, predominantly governed by exchange interactions and symmetry-breaking mechanisms. This unique feature distinguishes the altermagnetic system from conventional topological materials, which typically require strong SOC or external fields to induce band splitting. The weak dependence on SOC suggests that the system can achieve significantly longer spin relaxation times ($\tau_s$) due to reduced spin-flip scattering, as $\tau_s$ is inversely proportional to the square of SOC strength. Additionally, the SOC-independent band splitting enhances temperature stability, as the spin-polarized states are less susceptible to thermal fluctuations that typically degrade SOC-driven effects. These advantages make the system a promising candidate for high-performance spintronic devices operating at finite temperature.

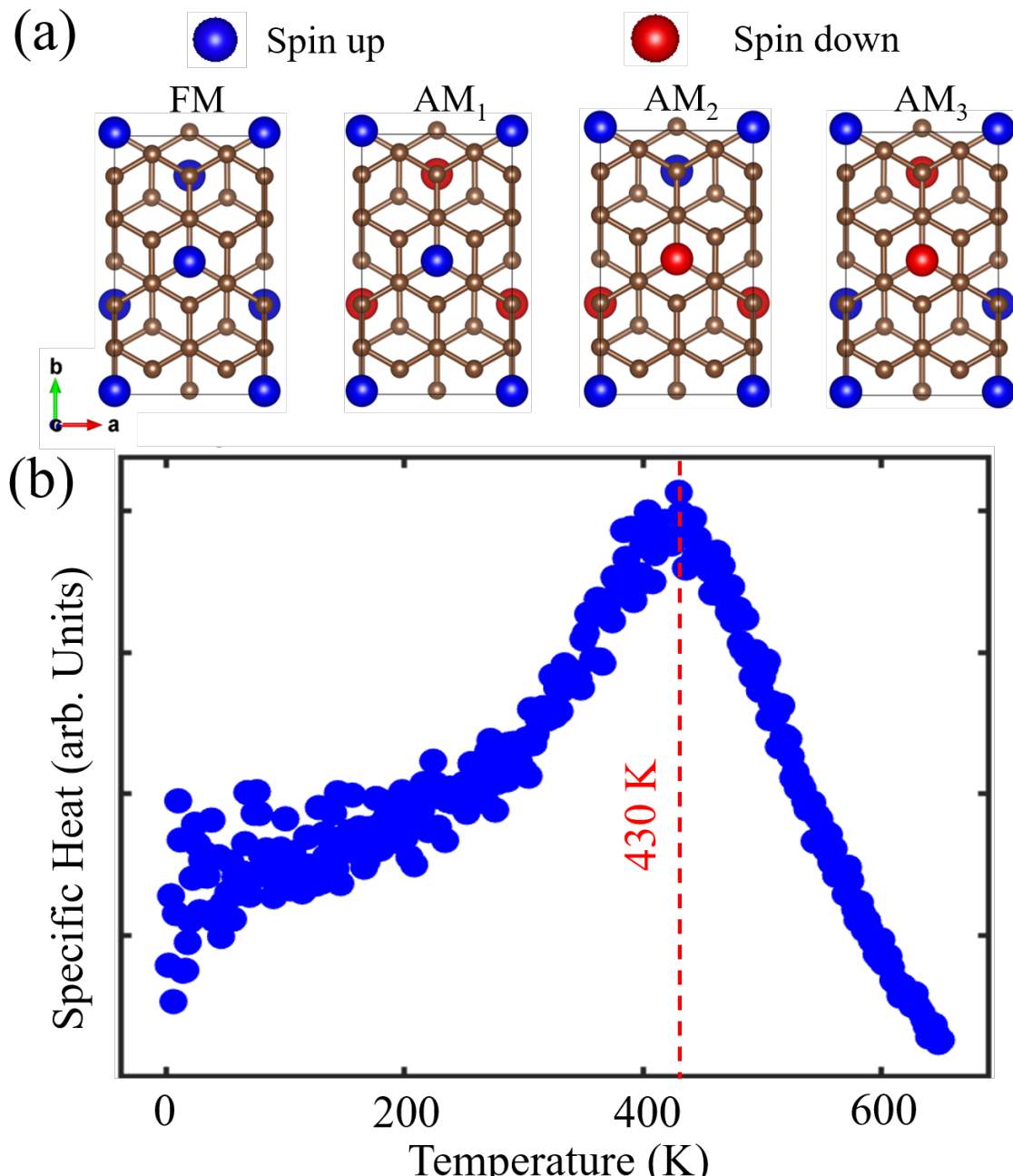

Fig 4. (a) Schematic representation of the magnetic exchange interaction of V-GIC supercells with hexagonal structure. Including nearest and next-nearest neighbor spin exchanges. The brown spheres represent carbon atoms, and the red and blue spheres represent vanadium atoms with spin up and spin down, respectively. (b) Specific heat versus temperature from Monte Carlo simulations, transition temperature is estimated as ~430K.

First-principles calculations incorporating SOC reveal a moderate magnetic anisotropy energy (MAE) of 0.6 meV per formula unit, with the easy axis aligned along the crystallographic [100] direction (Fig. 1b). This MAE magnitude, comparable to benchmark 2D $VI_3$ (0.8 meV) and $CrI_3$ (0.6 meV) [58] , suggesting comparable spin-orbital locking strength. To account for the effects of finite anisotropy and avoid overestimating the $T_m$, we performed Monte Carlo simulations based on the Heisenberg model. The four supercell configurations (Fig. 4a) exhibit distinct energetic hierarchies: FM, $AM_1$ (ground state), $AM_2$, and $AM_3$, with energy differentials relative to $AM_1$ measuring 0.54 eV, 0.27 eV, and 0.65 eV, respectively. The system can be effectively simplified to a hexagonal lattice, as illustrated in Figure 4(a), and described by an effective spin Hamiltonian that captures the magnetic interactions within the system.

$$\mathcal{H}_{\text{heisenberg}} = -2\text{J}_1 \sum_{\langle i,j \rangle} S_i \cdot S_j - 2\text{J}_2 \sum_{\langle\langle i,j \rangle\rangle} S_i \cdot S_j - 2\text{J}_3 \sum_{\langle\langle i,j \rangle\rangle} S_i \cdot S_j - \text{D} \sum_{i} |S_i^z|^2 \quad (2)$$

Here, $J_1$, $J_2$ and $J_3$ represent the isotropic exchange parameters for the first, second and third nearest neighbors, respectively, where ⟨ij⟩, ⟨⟨ij⟩⟩and ⟨⟨⟨ij⟩⟩⟩ nearest, second, and third neighbor spins. $D$ represents the single-ion anisotropy term, which accounts for the anisotropic magnetic interactions. Since the MAE and isotropic exchange parameters from the graphite is relatively weak, only the MAE and isotropic exchange parameters of V is considered in calculating, and the MAE of C atoms is assumed to be zero. The energy expressions for the four systems illustrated in Figure 4(a) can be written sequentially as follows:

$$E_{FM} = -E_0 - (3J_1 + 6J_2 + 3J_3)|\vec{S}|^2 - D|\vec{S}_z|^2$$
$$E_{AM1} = -E_0 - (-3J_1 + 6J_2 - 3J_3)|\vec{S}|^2 - D|\vec{S}_z|^2$$
$$E_{AM2} = -E_0 - (J_1 - 2J_2 - 3J_3)|\vec{S}|^2 - D|\vec{S}_z|^2$$
$$E_{AM3} = -E_0 - (-J_1 - 2J_2 + 3J_3)|\vec{S}|^2 - D|\vec{S}_z|^2 \quad (\mathbf{3})$$

The calculated exchange interaction parameters are determined to be $J_1 = -16.993$ meV, $J_2 = 3.466$ meV, $J_3 = 3.713$ meV and $D = 0.315$ meV. The specific heat curves in Figure 4(b) show that the $T_m$ for V-GIC is ~430 K. This prediction indicates that V-GIC are materials that can remain stable at temperatures well above room temperature and exhibit alternating magnetism. Its metallic properties and high phase transition temperature also provide broad prospects for its subsequent practical application.

## CONCLUSION

In summary, this work establishes V-GIC ($C_{16}V_2$) as a prototypical altermagnet, leveraging symmetry engineering to realize a compensated magnetic phase with momentum-dependent spin polarization. Combining first-principles calculations with symmetry analysis, we demonstrate that the hexagonal crystal structure ($P6_3mc$) and antiferromagnetic ordering of vanadium atoms generate a spin space group ($P^{-1}6^{1}_{3}m^{-1}c^{\infty m}1$) containing the key operation $[C_2||C_{6z}|t]$, which enforces alternating spin-split band structures while preserving zero net magnetization. Hallmark of the altermagnetic phase—robust spin splitting (~272 meV along $M_1'$–Γ–$M_1$) and spin-degeneracy along symmetry-protected directions, originate fundamentally from the non-relativistic spin group symmetry. Crucially, the spin splitting exhibits negligible dependence on spin-orbit coupling, underscoring the dominance of exchange interactions over relativistic effects and suggesting enhanced spin transport properties with prolonged spin relaxation times. Monte Carlo simulations further predict a magnetic transition temperature ($T_m$) of ~430 K, positioning this system among the rare class of altermagnets capable of sustaining magnetic order above room temperatures. The synergy between symmetry-protected spin structures, SOC-independent splitting, and elevated $T_m$ highlights the unique advantages of V-GIC for practical spintronic applications, including zero-field spin-polarized current generation and topologically robust spin manipulation. These findings not only advance intercalated graphite as a tunable platform for exploring altermagnetism but also provide a blueprint for designing symmetry-driven quantum materials. Future efforts could extend this approach to other transition metal-graphite systems, optimize interfacial coupling in heterostructures, and explore crossover phenomena between altermagnetism and topological states.

## ACKNOWLEDGMENTS

This work was supported by the National Key Research and Development Program of China (Grants No. 2024YFB3713803), the Key Program of the National Natural Science Foundation of China (Grants No. 52274386). This work was also supported by Shanghai Technical Service Center of Science and Engineering Computing, Shanghai University. We gratefully acknowledge HZWTECH for providing computation facilities.

## APPENDIX

Table S1. Symmetry analysis and magnetic properties of the altermagnetic ground state of V-GIC (*P6₃mc*, No. 186). The table lists the Wyckoff position, site symmetry, and local magnetic moment ($\mu_B$) for each atomic site.

| Atom | Site symmetry | Fractional Coordinates (*x*, *y*, *z*) | Magnetic moment ($\mu_B$) |
|---|---|---|---|
| $C_1$ (2a) | 3m. | (0, 0, 0.262) | -0.075 |
| $C_2$ (2a) | 3m. | (0, 0, 0.762) | 0.075 |
| $C_3$ (6c) | .m. | (0.330, 0.165, 0.253) | 0.039 |
| $C_4$ (6c) | .m. | (0.165, 0.330, 0.753) | -0.039 |
| $C_5$ (6c) | .m. | (0.500, -0.001, 0.246) | 0.007 |
| $C_6$ (6c) | .m. | (0.500, 0.001, 0.746) | -0.007 |
| $C_7$ (6c) | .m. | (0.835, 0.165, 0.253) | 0.039 |
| $C_8$ (2b) | 3m. | (0.667, 0.333, 0.761) | 0.059 |
| $C_9$ (6c) | .m. | (0.001, 0.500, 0.246) | 0.007 |
| $C_{10}$ (6c) | .m. | (-0.001, 0.500, 0.746) | -0.007 |
| $C_{11}$ (2b) | 3m. | (0.333, 0.667, 0.261) | -0.059 |
| $C_{12}$ (6c) | .m. | (0.165, 0.835, 0.753) | -0.039 |
| $C_{13}$ (6c) | .m. | (0.500, 0.500, 0.250) | 0.007 |
| $C_{14}$ (6c) | .m. | (0.500, 0.500, 0.750) | -0.007 |
| $C_{15}$ (6c) | .m. | (0.835, 0.670, 0.253) | 0.039 |
| $C_{16}$ (6c) | .m. | (0.670, 0.835, 0.753) | -0.039 |
| $V_1$ (2b) | 3m. | (0.333, 0.667, -0.021) | 1.766 |
| $V_2$ (2b) | 3m. | (0.667, 0.333, 0.479) | -1.766 |

Note S1. Symmetry analysis

I. space group symmetry

In this section, we detail the crystallographic symmetry of the system in its high temperature paramagnetic phase, which is chracterized by the non-centrosymmetric space group *P6₃mc*. The overarching goal of this symmetry analysis (section I, II and III) is to elucidate the relationships among the space group (SG), the magnetic space group (MSG) and the spin group (SSG) of V-GIC, with a particular focus on how these symmetries yield momentum-dependent band structure. The space group P6₃mc of our V-GIC comprises 12 symmetry operations, as listed using Seitz sysmbols in Table. S2. Among them, 1 and -1 denotes identity and inversion , respectively. The operation $6_{001}$ denotes a sixfold rotation around the [001] axis, with $6^+$ signifying a clockwise rotation and $6^-$ denoting counterclockwise rotation. $m_{100}$ represents a mirror reflection with the (100) plane as the mirror plane. Vectors such as $[00\frac{1}{2}]$ denote glide translation ($\tau$). Other operations follow similar conventions.

Table S2. Symmetry operations of the SG *P6₃mc* [1].

| | | | |
|---|---|---|---|
| $\{1\|0\}$ | : (*a*, *b*, *c*) →(*a*, *b*, *c*) | $\{m_{110}\|0\}$ | : (*a*, *b*, *c*) → (-*b*, -*a*, *c*) |
| $\{3_{001}^{+}\|0\}$ | : (*a*, *b*, *c*) → (-*b*, *a*-*b*, *c*) | $\{m_{100}\|0\}$ | : (*a*, *b*, *c*) → (-*a*+*b*, *b*, *c*) |
| $\{3_{001}^{-}\|0\}$ | : (*a*, *b*, *c*) → (-*a*+*b*, -a, *c*) | $\{m_{010}\|0\}$ | : (*a*, *b*, *c*) → (*a*, *a*-*b*, *c*) |
| $\{2_{001}\|00\frac{1}{2}\}$ | : (*a*, *b*, *c*) → (-*a*, -*b*, $c+\frac{1}{2}$) | $\{m_{1\bar{1}0}\|00\frac{1}{2}\}$ | : (*a*, *b*, *c*) → (*b*, *a*, $c+\frac{1}{2}$) |
| $\{6_{001}^{+}\|00\frac{1}{2}\}$ | : (*a*, *b*, *c*) → (*a*-*b*, *a*, $c+\frac{1}{2}$) | $\{m_{120}\|00\frac{1}{2}\}$ | : (*a*, *b*, *c*) → (*a*-*b*, -*b*, $c+\frac{1}{2}$) |
| $\{6_{001}^{-}\|00\frac{1}{2}\}$ | : (*a*, *b*, *c*) → (*b*, -*a*+*b*, $c+\frac{1}{2}$) | $\{m_{210}\|00\frac{1}{2}\}$ | : (*a*, *b*, *c*) → (-*a*, -*a*+*b*, $c+\frac{1}{2}$) |

II. Magnetic group symmetry

Considering the introduction of magnetism, we further explain the momentum dependence of the bands through magnetic space group. The MSG of our systems is identified as $P6_3'm'c$ (No. 186. 205). In MSG, the 12 symmetry operations of the parent group are partitioned into a unitary subgroup ($G_U$), and a coset of antiunitary operations $G_{AU}=\mathcal{T}(G- G_U)$, as listed in Table. S3. It is precisely these antiunitary operations gurantee the equivalence of high-symmetry k-points in the BZ, there by enabling the band structure to exhibit degeneracy.

Table S3. Symmetry operations of MSG $P6_3'm'c$.

| $G_U$ | | $G_{AU}$ | |
|---|---|---|---|
| $\{1\|0\}$ | : ($a$, $b$, $c$) →($a$, $b$, $c$) | $\{2'_{001}\|00\frac{1}{2}\}$ | : ($a$, $b$, $c$) → (-$a$, -$b$, $c$+$\frac{1}{2}$) |
| $\{3^+_{001}\|0\}$ | : ($a$, $b$, $c$) → (-$b$, $a$-$b$, $c$) | $\{6'^+_{001}\|00\frac{1}{2}\}$ | : ($a$, $b$, $c$) → ($a$-$b$, $a$, $c$+$\frac{1}{2}$) |
| $\{3^-_{001}\|0\}$ | : ($a$, $b$, $c$) → (-$a$+$b$, -a, $c$) | $\{6'^-_{001}\|00\frac{1}{2}\}$ | : ($a$, $b$, $c$) → ($b$, -$a$+$b$, $c$+$\frac{1}{2}$)) |
| $\{m_{1\bar{1}0}\|00\frac{1}{2}\}$ | : ($a$, $b$, $c$) → ($b$, $a$, $c$+$\frac{1}{2}$) | $\{m'_{1\bar{1}0}\|0\}$ | : ($a$, $b$, $c$) → ($b$, $a$, $c$+$\frac{1}{2}$) |
| $\{m_{120}\|00\frac{1}{2}\}$ | : ($a$, $b$, $c$) → ($a$-$b$, -$b$, $c$+$\frac{1}{2}$) | $\{m'_{100}\|0\}$ | : ($a$, $b$, $c$) → ($a$-$b$, -$b$, $c$+$\frac{1}{2}$) |
| $\{m_{210}\|00\frac{1}{2}\}$ | : ($a$, $b$, $c$) → (-$a$, -$a$+$b$, $c$+$\frac{1}{2}$) | $\{m'_{010}\|0\}$ | : ($a$, $b$, $c$) → (-$a$, -$a$+$b$, $c$+$\frac{1}{2}$) |

Considering $k_x$-$k_y$ plane, where $k_z$=0, we can represent the **k**-vector as ($u$, $v$, 0). Within the MSG, there exists a $G_{AU}$ that the **k**-vector remains unchanged ($u$=[-π/a, π/a] , $v$=[-π/b, π/b]):

$$\mathcal{T}\{2'_{001}|00\tfrac{1}{2}\} : (u, v, 0) \rightarrow (u, v, 0)$$

While $k_z$=π/c, the antiunitary operation keeps the $\vec{k}$-vector is:

$$\mathcal{T}\{2'_{001}|00\tfrac{1}{2}\} : \left(u, v, \frac{\pi}{c}\right) \rightarrow \left(u, v, -\frac{\pi}{c}\right) \equiv \left(u, v, \frac{\pi}{c}\right)\ (mod\ G)$$

The antiunitary operation $\mathcal{T}\{2'_{001}|00\frac{1}{2}\}$ connecting both spin sublattice flips the spin in these two planes, preserving the spin degeneracy in Γ-M-K-Γ and A-L-H-A as shown in Fig. 1 (a).

III. Spin group symmetry

While MSG are useful, a comlete understanding of momentum-dependent band splitting requires the use of non-realistivistic spin groups that decouple spatial and spin operations. A spin space group could be expressed as $G_{SS} = G_{NS} \times G_{SO}$, where $G_{NS}$ represents the non-trivial spin groups and $G_{SO}$ stands for spin-only group. For collinear systems, $G_{SO} = SO(2) \rtimes Z_2^K$(international notation is $^{\infty m}1$), where SO (2) =$\{\{E\|E\}, \{U_z(\emptyset)\|E\}\}, \emptyset \in [0,2\pi)$. For collinear FM materials, which only contains one type of spin lattice, $G_{NS}$=$\{E\|G_0\}$ and $G_0$ is one of the 230 SGs. In term of systems with two spin sublattice such as AFM, $G_{NS} = \{E\|L_0\} + \{\mathcal{T}\|AL_0\}$, $L_0$ is an index-two normal subgroup of the $G_0$ containing all spatial symmetry operations that do not exchange from different sublattices. $\mathcal{T}$ and $A$ represents spin reversal and the symmetry operations connecting two sublattices, respectively [2]. According to the difference of $A$, SSGs could be divided into 3 classifications. When $A$ represents the space inversion $\mathcal{P}$ or lattice translation $t$, the spin groups belong to the type-II and type-IV, respectively, both exhibiting zero net magnetization as AFM systems. While $A$ represents a mirror reflection $M$, a rotation $C_n$ or other composite rotation operations, the spin group belonging to type-III represents the AM we are concerned with [3].

The V-GIC in our work has a hexagonal structure with in-plane collinear and zero-net antiferromagnetic interlayer coupling. The crystallograohic space group is $G_0$= P6₃mc , while the subgroup that preserves the sublattice is $L_0$=P3m1 (NO. 156), belonging to the $t$-type normal subgroup with the indices $i_t$=2, $i_k$=1. The symmetry operations $A$ connecting the sublattices with antiparrell magnetic directions are $\left\{6_{001}\middle|00\frac{1}{2}\right\}$, $\left\{2_{001}\middle|00\frac{1}{2}\right\}$, $\left\{m_{210}\middle|00\frac{1}{2}\right\}$, $\left\{m_{120}\middle|00\frac{1}{2}\right\}$ and $\left\{m_{1\bar{1}0}\middle|00\frac{1}{2}\right\}$ (corresponding to $C_{tz}t$, $C_{2z}t$, $M_{210}t$, etc. ) . The nontrival spin group can be constructed from the coset decompostion $G_{NS} = \{E\|L_0\} + \{\mathcal{T}\|AL_0\}$:

$$G_{NS} = \{E\|L_0\} \cup \left\{\bar{1}\|6_{001}\middle|00\frac{1}{2}\right\} L_0 \cup \left\{\bar{1}\|2_{001}\middle|00\frac{1}{2}\right\} L_0 \cup \left\{\bar{1}\|m_{210}\middle|00\frac{1}{2}\right\} L_0 \cup \left\{\bar{1}\|m_{120}\middle|00\frac{1}{2}\right\} L_0 \cup \left\{\bar{1}\|m_{1\bar{1}0}\middle|00\frac{1}{2}\right\} L_0,$$

the corresponding international notation is $P^{-1}6_3^1m^{-1}c$. Considering the spin-only group $G_{SO}$, full spin space group is $P^{-1}6_3^1m^{-1}c^{\infty m}1$, belonging to the type-III class of spin groups that defines AM.

We now analyzethe unique momentum dependence of the energy bands using these spin group operations. The **k**-vector is denotes as ($u$, $v$, $w$). ($u$=[-π/a, π/a] , $v$=[-π/b, π/b], $w$=[-π/c, π/c]).

First, the spin group symmetries impose several spin-degenerate nodal planes in the BZ. For the vertical glide planes, such as $u$=$v$, the operation $\left\{m_{1\bar{1}0}\middle|00\frac{1}{2}\right\}$ leaves **k**-vector invariant:

$$\left\{m_{1\bar{1}0}\middle|00\frac{1}{2}\right\}: (u, u, w) \rightarrow (u, u, w).$$

The spin group symmetry $[C_2||M_{1\bar{1}0}t]$ satisfies:

$$\epsilon(s, \mathbf{k}) = [C_2||M_{1\bar{1}0}t]\epsilon(s, \mathbf{k}) = \epsilon(-s, M_{1\bar{1}0}t\mathbf{k}) = \epsilon(-s, \mathbf{k}).$$

As for $v$=0 and u=0,

$$\left\{m_{120}\middle|00\frac{1}{2}\right\}: (u, 0, w) \rightarrow (u, 0, w)$$

$$\epsilon(s,\mathbf{k}) = [C_2||M_{120}t]\epsilon(s,\mathbf{k}) = \epsilon(-s,M_{120}t\mathbf{k}) = \epsilon(-s,\mathbf{k})$$
$$\left\{m_{210}\middle|00\tfrac{1}{2}\right\}: (0,v,w) \rightarrow (0,v,w)$$
$$\epsilon(s,\mathbf{k}) = [C_2||M_{210}t]\epsilon(s,\mathbf{k}) = \epsilon(-s,M_{210}t\mathbf{k}) = \epsilon(-s,\mathbf{k})$$

This relationship holds for the three vertical planes (the green planes in Fig. S1), proving they are spin-degenerate. Similarly, for the horizontal planes at *w*=0 and w=π/c, the screw axis operation $\left\{2_{001}\middle|00\tfrac{1}{2}\right\}$ maps a vector **k** to **-k.**

$$\left\{2_{001}\middle|00\tfrac{1}{2}\right\}: (u,v,0) \rightarrow (-u,-v,0)$$
$$\left\{2_{001}\middle|00\frac{1}{2}\right\}: \left(u,v,\frac{\pi}{c}\right) \rightarrow \left(-u,-v,\frac{\pi}{c}\right) \equiv \left(-u,-v,-\frac{\pi}{c}\right)\ (mod\ G)$$

Combined with the general inversion symmetry $[\bar{C}_2\|\mathcal{T}]$ in collinear magnets, $\epsilon(-s,-\mathbf{k}) = [\bar{C}_2\|\mathcal{T}]\epsilon(-s,-\mathbf{k}) = \epsilon(-s,\mathbf{k})$, this confirms the blue planes in Fig. S1 are also spin-degenerate.

Away from these nodal planes, for a general **k**-point, the rotational nature of the connecting operations leads to a robust spin splitting.

$$\left\{6^{+}_{001}\middle|00\tfrac{1}{2}\right\}: (u,v,w) \rightarrow (u-v,u,w)$$
$$\left\{6^{-}_{001}\middle|00\tfrac{1}{2}\right\}: (u,v,w) \rightarrow (v,v-u,w)$$
$$\epsilon(s,\mathbf{k}) = [C_2||C_{6z}t]\epsilon(s,\mathbf{k}) = \epsilon(-s,C_{6z}t\mathbf{k})$$

Since $\mathbf{k} \neq C_{6z}t\mathbf{k}$, it follows that $\epsilon(s,\mathbf{k}) \neq \epsilon(-s,-\mathbf{k})$. This demonstrates a non-zero spin polarization($\boldsymbol{P}_s$≠0) at general momenta. Moreover, due to the transformation connects opposite spins at different **k**-points, the spin polarization necessarily alternates across the Brillouin zone, which is the defining characteristic of the altermagnetic band structure.

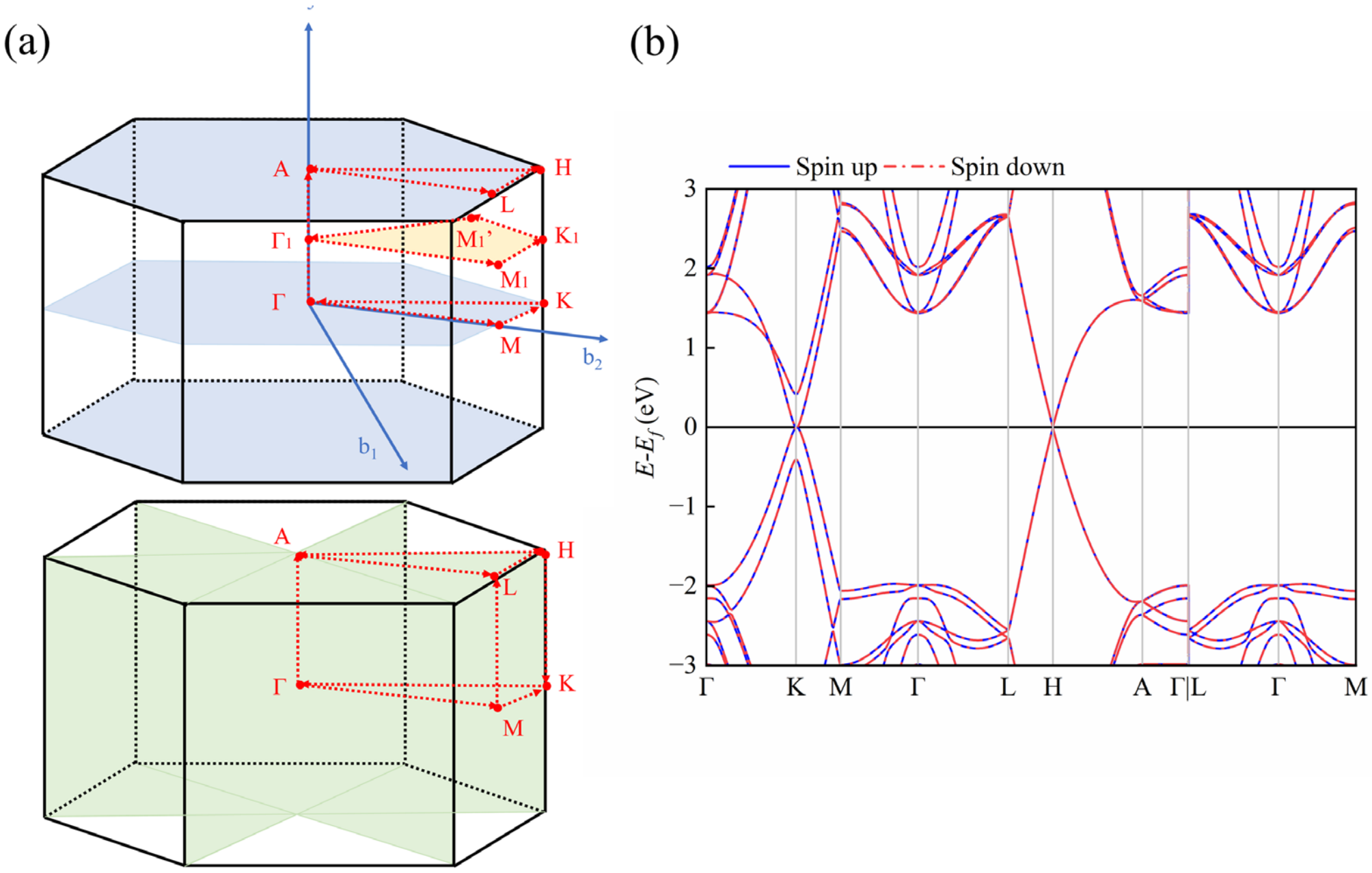

Fig. S1 (a) The first Brillouin zone of the V-GIC, along with the high-symmetry points path used in the band structure calculations. (b) Band structure of the pristine graphene, shown for reference. Our result is excellent agreement with previous work [4].

---